\documentclass[preprint, a4paper,
showpacs,showkeys,groupedaddress]{revtex4}
\usepackage{amsmath,amsfonts,amssymb,epsfig,epstopdf,url,array} 
\usepackage{hyperref}
\hypersetup{
linkcolor=blue,
citecolor=red,
colorlinks=true,
pdfborder={0 0 0},
}

\def\Op{{\rm O}} 
\def\derpar#1#2{\frac{\partial{#1}}{\partial{#2}}} 
\def\Tan{{\rm T}} 
\def\P{{\cal P}} 
\def\beq{\begin{equation}} 
\def\eeq{\end{equation}} 
\def\d{{\rm d}} 
\def\Complex{\mbox{\it C}} 
\def\sta{|\psi \rangle }
\def\p{ \partial}

\newtheorem{definition}{Definition}
\newtheorem{teor}{Theorem}[section] 
\newtheorem{lemma}{Lemma}
\newtheorem{prop}{Proposition}[section] 
\newtheorem{corol}{Corollary} 
\newtheorem{rem}{Remark}[section] 
\newtheorem{obs}{Observation}[section] 
\newtheorem{example}{Example}

\begin{document} 

\def\bea{\begin{eqnarray}} 
\def\eea{\end{eqnarray}} 

\floatsep 30pt 
\intextsep 30pt 

\title{Remarks on the geometric quantization of a class of harmonic oscillator type potentials} 

\author{Felix Iacob} 

\email {felix.iacob@gmail.com }

\affiliation{ West University of Timi\c soara,\\ 
300223 V. P\^ arvan 4, Timi\c soara, Romania.\\ 
} 
\begin{abstract} 
The conditions that must be fulfilled by a certain physical system to apply geometric quantization prescription on it are investigated. 
These terms are sought as mathematical requirements, which can be traced in an analysis of integrable systems, from the perspective of both potential function and Hamiltonian vector field.
The answer is found in momentum map critical points. 
Basically, a certain disposal of points that allow a momentum map $C^*$ isomorphism, of observables, with harmonic oscillator momentum map enforce geometric quantization rules. 

Following the general theory, two newly presented examples, which exhibits these properties, are quantified through geometric quantization prescription. The Lennard-Jones' type potential is one of the examples, it is known as describing molecules in interaction. We end with a third example that shows the local isomorphism of potentials do not induce $C^*$ isomorphism of observables.
\end{abstract} 


\maketitle 

\section{Introduction}\label{intro} 
Harmonic oscillator (HO) is a successful model regarding its quantization, both canonical and geometrical one, being an example of solvability. 
The theoretical study of HO occupy a central place in quantum mechanics, one can find a range of its applications in the real world of atomic and molecular systems. 
The potential function is the key element, that defines the evolution of a system.
Only some functions have proven to be mathematically reliable and fewer meet the conditions to describe a physical system. 
Exactly solvable potentials, in quantum mechanics, have been proposed starting with the early mid of the last century and continue to be an interesting topic. 
Beside hydrogen atom, the other fundamental quantum mechanic example of solvable system, there are other solvable physical systems, described by potentials, known as: the Morse potential \cite{2}, the Eckart potential \cite{3}, the Rosen-Morse potential \cite{4}, the trigonometric and hyperbolic Pöschl-Teller potentials \cite{5}, the Manning-Rosen potential \cite{6}, the Woods-Saxon potential \cite{7}, the Scarf potential \cite{8} and the pseudo-Gaussian potential \cite{pgo}.

In this paper, we introduce a class of basic functions which define solvable potentials, for certain physical systems, that can be reduced to the HO potential by some elementary transformations and, as a matter of fact, admits geometric quantization (GQ) procedure. 
The study is made upon the geometrical properties of phase space of physical systems from the perspective of both the Hamiltonian functions, this is the foliation perspective, and the flow of the vector fields, this is the dynamical system perspective. 
Mainly, critical points of the system, known as singularities, are investigated in order to find the conditions of integrability and thereby the dynamics of the system.
Our reasoning is based on famous results of both Morse lemma, regarding the study of critical points of Hamiltonian function, and Eliasson theorem, which gives the integrability conditions.
We conclude by a proposition that gives the sufficient conditions needed by a physical system in order to be integrable in HO sense, this means to apply GQ procedure (given by Simms \cite{Simm}) used to the quantization of energy states of HO.

In our approach, we make use of GQ because, as it is known, it is a mathematically rigorous global generalization of the canonical quantization (CQ) technique. 
As a matter of fact GQ is suited for our approach, which is a generalization for an entire class, because there is no need for a specific physical system to be pointed out. 
Having this in mind, we start with a general form for potential, namely $V$, showing which are the requirements that have to be accomplished by it in order to become the HO potential. 
The classical systems are described using the technique of symplectic manifolds, which provides the adequate Hamiltonian formulation of autonomous mechanics (the time-dependent systems are not considered in this work). 
When GQ, based on symplectic manifolds, takes into account the phase space geometry of the classical physical system and aims to find a quantum counterpart, one can say that GQ follows the so called Dirac program \cite{AM}. 
Methods of GQ have been applied with great success to the theory of representation of Lie groups \cite{Kon}, however, its usefulness in applications to quantum theory has been rather limited. 
At least in this terms the cornerstones of quantum mechanics, the harmonic oscillator and the hydrogen atom, GQ gives results in agreement with those of CQ \cite{K,AK,S,Simm,Kumm,Cord}. 
The geometric quantization achievements are still under those accomplished by canonical quantization. In this view, every new example contributes to the edification of GQ. 
Being accepted at this level, GQ can further show its usefulness when one needs to take into account the topological and geometrical structure of the classical phase space, for instance in the case of Gravitation \cite{Gr}. 

Particular systems that can be reduced to the quantization scheme of HO have already been reported, we notice here the nonisotropic harmonic oscillator \cite{MM}, the rigid body with a single rotor about the third principal axis and an internal torque \cite{Puta}, the Kepler problem obtained from HO using a Segre map \cite{GS} and the symmetric rigid body \cite{I}. 

In quantum mechanics finding new solvable potentials is a question of central interest and a subject of long time studies. We know only a handful of classical potentials for which we can determine explicitly all the bound states in terms of elementary transcendental functions and to write explicitly their energies in terms of quantum numbers. 
We hope that contribution of this work to GQ theory illustrates the general theory with new specific examples and with the same extent brings out the properties required by the potential of certain physical systems in order to be integrable by admitting GQ procedure. 
In quantum physics the Schr\" odinger picture (or representation) is a formulation where the physical states evolve in time. Opposite, we can think the states remain constant and the operators (observables) evolve in time. Physically this is the difference between active and passive viewpoint. In GQ the representation is given by polarization. As it is usual, we call Schr\" odinger representation that one were the polarization is spanned by the vector field determined by position coordinate. The dynamics of the system is given by Schr\" odinger equation (SE), a second order partial differential equation, which describes the evolution of the states. The potential function is part of the differential operator, acting on Hilbert space, called Hamiltonian. 
An extensive study on Schr\" odinger equations reducible to other hypergeometric or rescaled confluent equation can be consulted in ref. \cite{fluegge, AHP, milson}.

The rest of this paper is organized as follows:
In the first section we present some aspects of both GQ and CQ focusing on what it is considered an achievement in each of them. The GQ steps are mentioned along with Schr\" odinger representation as it comes naturally from phase space of the physical system.
In the second section, the behavior of physical system is studied through its critical points, also known as stationary points, which define how states evolve. The study is made from the perspective of both potential function (passive mode) and Hamiltonian vector field (dynamic mode), thus we split this section in two subsections. In the former, with the help of Morse lemma, we seek mathematical aspects of a class of potentials that can be reduced to harmonic oscillator one. 
In the latter, making use of Eliasson theorem, we seek mathematical conditions that have to be fulfilled by Hamiltonian vector field such that the system is integrable in HO sense. In the end of the latter subsection we give a result that states about the conditions to be fulfilled by the physical system to be integrable with the momentum map isomorphic with the HO momentum map, from both perspective.
The next section is dedicated to geometric quantization of HO in holomorphic representation using a particular K\"ahler polarization. We emphasize the importance of choosing the correct Hamiltonian acting on holomorphic sections as a correspondent of energy function of the system. Here one should consider the commutation relation of complex coordinates, as it comes from quantum mechanic in the energy or H-representation.
In the last section we introduce two physical systems that illustrate the reasoning exposed throughout the paper. We show that the potentials belong to the introduced class of functions and therefore GQ prescription exposed in the previous section can be applied.
Finally, we want to emphasize the importance of the conditions, developed here, by presenting an example of a physical system that has the potential local isomorphic with HO, but does not admit GQ having no isomorphism between momentum maps. 
The paper ends with conclusions.

\section{About geometric quantization.} 

An achievement in quantum mechanics is to build the spectrum and the corresponding eigenfunctions for energy operator, using Schr\" odinger's equation with some given boundary conditions. Once this is done, we can say that we have an integrable physical system, further on this subject is exposed in \cite{Fordy}. 
In geometric quantization procedure, for a given physical system, the quantization rules are used to obtain a Hilbert space of states and a set of operators acting on Hilbert space, representing quantum observables. 
Formally, we can say that the GQ procedure involves three major steps: prequantization, polarization, and metaplectic correction. 
Prequantization gives a preliminary Hilbert space and a complete, but reducible, representation of the classical observables. 
Prequantization is not enough to get the correct energy spectrum of physical systems, thus some additional structures where introduced to obtain a quantization of a symplectic manifold. One of these is a polarization, and this generally leads to a rather severe technical complications. 
The polarization is needed to reduce the pre-Hilbert space, this is made by preserving the complete commuting set of observables. 
The complication occurs due to lack of natural measure on the space of quantum states in most cases, further more when such measure exists sometimes GQ is still not completely correct regarding the spectrum of energies. 
In this case it is needed to modify the quantization scheme to what is known as half-form or metaplectic quantization. At this stage GQ becomes successful, but the reverse of the medal is it becomes a little bit complicated and unwieldy.
The introduction of a metaplectic structure provides the measure in terms of which the quantum Hilbert space inner product is defined. 

Physical systems admit the Schr\" odinger representation, which comes naturally from phase space, by introducing the cotangent bundle $M =T^*Q$, with canonical basis $\{q_i,p_j\}$ and standard symplectic form $\Omega = dq_i \wedge dp_i$. 
The polarization $P$ is taken to be the span of the vertical vector fields $\left\{ \frac{\partial}{\partial p_i}\right\}_{i=1\ldots n}$. 
The polarized sections $\psi$ are the sections for which $\frac{\partial\psi}{\partial p_i} =0$, i.e. those which are constant along the fibers. In this way the dimension of pre-Hilbert space is reduced, the sections depend on half of variables i.e. $\psi=\psi(q^i)$ and form the Hilbert space. 
The corresponding quantum operators for the momenta respectively position are written as: 
\beq \label{oppq}
\Op_{p_j} = -i\hbar \derpar{}{q^j} \ ; \ 
\Op_{q^j} = q^j.
\eeq 
Physics sets Hamilton's function to be the most significant observable, having a kinetic and potential part, $H=K+V$. It defines the energy and describes the motion of system.
The dynamics is generated by a smooth potential $V(q)$ acting on Hilbert space, described with the help of equation of motion: 
\begin{prop} \label{ham} 
The Hamiltonian function $H = p^2+V(q)$, up to some factors and constants, produces a vector field, $X_{H}$, such that the dynamics is given by Hamilton's equations $i_{X_{H}}\Omega=-dH$. 
\end{prop} 
\noindent The classical observable $H$ gives rise to the Hamiltonian vector field $X_{H}$, the flow of $X_{H}$ is called the Hamilton flow. 

\section{Critical points.}
\subsection{Critical points of Hamiltonian function.}\label{cph}

It is important for the dynamics of the system to know how critical points of the Hamiltonian function are managed. 
We are focusing around the critical points of the potential, which tell us about the change in shape of the potential. This information can be used to make a qualitative prediction about the spectra of Hamiltonian operator. 

Our reasoning is to show, by Morse lemma, the existence of a diffeomorphism in a neighborhood of a nondegenerate singular point of a smooth function that takes the given function to its quadratic part. In this way, we can reduce our system to one having quadratic potential. 
With other words, if the potential admits critical points, by Morse lemma, the topological changes of manifold are put into one-to-one correspondence with these points. 
Let us see how this things work, we begin with some needful definitions, briefly presented, but an exhaustive presentation can be found in \cite{Arnold-cp}.

\begin{definition} \hfill
\begin{itemize} 
\item The critical points of a smooth function are the points where the differential vanishes. 
\item A critical point is nondegenerate if the second differential (Hessian) is a nondegenerate quadratic form. 
\item The index $\lambda$
of nondegenerate critical point is the dimension of maximal subspace on which the Hessian is negative definite. 
\item A smooth function is called Morse function if all critical points are nondegenerate. 
\end{itemize} 
\end{definition} 
\begin{lemma}[Morse] 
Let $f$ be a Morse function and $p\in M$ a critical point of manifold $M$. There are local coordinates $\{x_1,\ldots, x_n\}$, with $p=(0,\ldots,0)$ such that $f(p\prime)= f(p) -x^2_1,\ldots, -x^2_\lambda + x^2_{\lambda+1},\ldots,+x^2_n$ for every point $ p\prime \in U$ a small neighborhood, and $\lambda$ is the index of $f$ at $p$. 
\end{lemma} 
What we can say, as an immediate result of Morse lemma, for our considered physical system is that:
\begin{corol} \label{morsepot} 
Let potential $V(q)$ be a Morse function, then in some neighborhood of a critical point the potential can be represented in the Morse formulation: 
\beq 
V(q)= \pm q^2_1 \pm \ldots \pm q^2_n 
\eeq 
this is made using canonical coordinates. 
\end{corol} 
It is easy to see that for a Morse function the critical points are isolated. Hence, the sets of critical points is a 0 - dimensional manifold. 
In physics the Morse function critical points are called stationary points, which can be stable or unstable. Of course, for physical reasons we are interested in the stable ones. 
\begin{definition} 
A stable stationary point is a critical point of a Morse function having the Hessian positive definite. 
\end{definition} 
With this we can state that the dynamics of physical system is locally generated by potential having the following form: 
\begin{prop} \label{potHO}
Let potential $V(q)$ be a Morse function, then in some neighborhood of a stable stationary point the potential can be represented in the Morse formulation: 
\beq 
V(q)= q^2_1 + \ldots + q^2_n .
\eeq 
\end{prop} 

We have seen that by studying the potential, we determine the conditions needed to write it locally as its quadratic part, the global dynamics have to be determined for each particular physical system. 

\subsection{Critical points of Hamiltonian flow}

The study of the singularities of Hamiltonian systems can be made by using two different approaches: one can study the Hamiltonian functions themselves, this is the foliation perspective (as we presented in subsection \ref{cph}), or one can analyze the flow of the vector fields, this is the dynamical viewpoint and follows in this section. 

\begin{definition} 
\begin{itemize} 
\item We say $X_{H}$ is complete if it generates a global flow on manifold. 
\item A function is called proper function if it has the property $f^{-1}(compact)=(compact)$. 
\end{itemize} 
\end{definition} 
 
\begin{lemma}[Gordon, \cite{Gordon}] 
Let $X$ be a ${\cal C}^1$ vector field on a manifold $M$ of class ${\cal C}^1$. Then $X$ is complete if there exist a ${\cal C}^1$-function $E$, a proper ${\cal C}^0$-function $f$, and constants $\alpha,\, \beta$, such that for all points in manifold $x\in M$ 
\begin{enumerate} 
\item $| X(E(x)) | \leq\alpha |E(x)| $ 
\item $| f(x) | \leq\beta |E(x)| $ 
\end{enumerate} 
\end{lemma} 
The above result, in the case of symplectic manifold, can be rewritten as follows: 
\begin{prop}[\cite{puta-pp} ] 
Let $(M, \Omega)$ be a smooth $2n$-dimensional manifold and $H\in \cal C^\infty$ the Hamilton's function on manifold. If $H$ is proper and bounded below then $X_{H}$ is complete. 
\end{prop} 
\begin{definition} 
The flow of a Hamiltonian $H$ on a $2n$-dimensional symplectic manifold $(M, \Omega)$ is said to be integrable, or completely integrable if there exist $n$ everywhere independent integrals $\{f_1=H, f_2, \ldots , f_n \}$ of the flow which are in involution. 
\end{definition}  

\begin{obs} \label{pb}
The algebra of classical observables, including Hamiltonians, comes naturally endowed with the Poisson bracket: $\{\cdot, H\}=\Omega(\cdot , X_{H})$. 
\end{obs}
The evolution of a function $f$ under the flow of $X_{H}$ is given by the equation $\dot f=\{H,f\}$. 
An integral of the Hamiltonian $H$ is a function which is invariant under the flow of $X_{H}$, i.e. a function $f$ such that $\{H, f\}= 0$. 
The functions $\{f_1,\ldots f_n\}$ are denoted by $F= (f_1,\ldots f_n)$, which is usually called the moment map. 

\begin{rem} 
On a $2n$-dimensional symplectic manifold, a completely integrable system is a moment map function. 
\end{rem} 
The moment map describe and takes all characteristics of the system. The singularity becomes a feature of the moment map. Accordingly, we seek systems with moment map isomorphic with that of HO by studying the singularities.
Singularities corresponding to fixed critical points of relative equilibrium of the system are one of the main characteristics of the dynamics. 
The Hessian in these points is a non-degenerate quadratic form. 
In accordance with the linear classification of Cartan subalgebras of $sp(2n,\mathbb{R})$, any such Cartan subalgebra has a basis build with three type of blocks: two uni-dimensional ones (the elliptic and the real hyperbolic) and a two-dimensional one called focus-focus \cite{Will}. 
\begin{rem}\label{ff}
There are only two types of non-degenerate singularities for integrable systems in dimension 2: hyperbolic (when the Hessian is indefinite) or elliptic (when the Hessian is positive or negative definite) 
\end{rem} 
 
Here is a major result, given by Eliasson \cite{Eliasson}, that gives a connection between integrable system and critical points:
\begin{teor}[Eliasson] 
The non-degenerate critical points of a completely integrable system are linearizable. 
\end{teor} 
  
This problem of symplectic linearization, closely related to the spirit of Morse lemma, was solved successfully by Vey and Colin de Verdiere \cite{Vey}. 
\begin{teor}[Vey] 
Let $F : (M^2,\Omega) \to \mathbb{R}$ be a function and let p be a non-degenerate singular point of $F$. Let $Q$ be the quadratic form corresponding to the Hessian of $F$ at $p$. 
Then there exists a local diffeomorphism from a neighborhood $Z$ of $p$ to a neighborhood of $0 \in \mathbb{R}^2$ taking $F$ to a function $\phi(Q)$. If the hessian $Q$ is positive definite the germ of the function $\phi$ characterizes the pair $(F, \,\Omega)$. If $Q$ is not definite then the jet at the point $p$ of the function $\phi$ characterizes the pair $(F, \,\Omega)$. 
\end{teor} 
\begin{rem} \label{vecHO}
As a consequence of this theorem, after putting $Q$ in coordinates form ($\Omega = \d x \wedge d y$) we can assume from now on that the foliation in a neighborhood of a singular point $p$ corresponding to $0$ is given by the vector field: 
\begin{itemize} 
\item $Y = - y\frac{\partial}{\partial x} + x \frac{\partial}{\partial y}$ when $ Q = x^2+y^2$ p is elliptic 
\item $Y = x\frac{\partial}{\partial x} - y \frac{\partial}{\partial y}$ when $ Q = xy$ p is hyperbolic 
\end{itemize} 
\end{rem} 
\noindent It is usual to call these ($x,y$) coordinates - Eliasson coordinates. 

The following result gives the conditions that have to be performed by a physical system such that it can be reduced to HO. This means an isomorphism of momentum maps, it relates the potential to Hamiltonian flow and its critical points of of each of two systems. If the isomorphism is established the GQ can be applied to it using set of rules to be followed in GQ of HO. 
We choose to give this result in two dimensions to avoid the signature discussions of Hessian. This means that, in accordance with the remark \ref{ff}, we do not consider the focus-focus case. 
Two dimensions in phase space $(q, p)$ represent a system with one degree of freedom in physical space, but the reasoning, of the following theorem, can be extended to the n-dimensional case $(q^n, p^n)$, leaving out the focus-focus case that is not relevant in applications in physics.

\begin{teor} \label{HOcond}
Let be a physical system acting on a 2-dimensional manifold, observed in Eliasson coordinates and described by its moment map. The system is integrable having the moment map isomorphic with HO moment map if at least one of the following conditions are fulfilled: 
\renewcommand{\theenumi}{\roman{enumi}}%
\begin{enumerate} 
\item The Hamiltonian is proper and bounded below. 
\item The Hamiltonian vector field is complete and has elliptic critical point. 
\end{enumerate} 
\end{teor} 
\begin{dem}
\renewcommand{\theenumi}{\roman{enumi}}%
In Eliasson coordinates, $(q, p)$ we have:
\begin{enumerate} 
\item The Hamiltonian is proper and bounded below implies there exists a stationary point such that the Hessian is positive definite, i.e. we have a stable stationary point. This Hamiltonian is a Morse function so according to prop. (\ref{potHO}), it takes the form $H= \Op_{p_j}^2 + \Op_{q^j}^2 $, which is the HO Hamiltonian, so the system is HO integrable. 
\item The Hamiltonian vector field is complete and has one or more elliptic critical points, according to rem. (\ref{vecHO}) it takes the form $H= \Op_{p_j}^2 + \Op_{q^j}^2 $, so the system is HO integrable. 
\end{enumerate} 
\end{dem}
\begin{obs}
Theorem \ref{HOcond} gathers together the foliation and flow of the vector fields perspective.
\end{obs}
\begin{obs}
In quantum mechanics, one typically describes a physical system with a C*-algebra of physical observables. The isomorphism in theor. \ref{HOcond} is meant as a $C^*$-algebra isomorphism, the unbounded operators are considered in the Weyl form \cite{Strocc}, in order to satisfy the canonical commutation relations.
\end{obs}
We are going to extend the above result through the following remark:
 
\begin{rem}\label{remmea}
In the hypothesis of theorem \ref{HOcond}, the equivalence of following conditions holds:
\begin{enumerate} 
\renewcommand{\theenumi}{{\em\roman{enumi}}}%
\item The system is integrable having the moment map isomorphic with HO moment map.
\item The Hamiltonian is proper and bounded below. 
\item The Hamiltonian vector field is complete and has elliptic critical points. 
\end{enumerate} 
\end{rem} 

\section{Harmonic oscillator quantization in Bargmann-Fock representation.}\label{BFR}

The representation is determined by the choice of polarization. The Schr\" odinger representation is given by a polarization spanned by the vector field determined by position coordinate. In the momentum representation the polarization is spanned by the vector field of momentum coordinate. The kernel between these representations is given the by Fourier transform.
The holomorphic or Bargmann-Fock representation is obtained when we use a particular K\"ahler polarization, in which we have to make a coordinate change, from the standard phase space coordinates $(p,q)$, by introducing the complex ones $\{ z_j, \bar z_j \}$ where $z_j := p_j+iq^j$. Quantum mechanic names this as energy or H-representation.
CQ method applies in both Schr\" odinger and Bargmann-Fock representation, while GQ method applies just in the last and its main features are presented below.
In this representation the symplectic form becomes: 
$$ 
\Omega=\frac{i}{2} \d \bar z_j \wedge \d z_j, 
$$ 
the structure $(\Tan^*Q,\Omega ,{\cal J})$ is a K\"ahler manifold, 
where the complex structure is given by: 
$$ 
{\cal J}\left(\derpar{}{p_j} \right) = \left(\derpar{}{q^j}\right) 
\quad , \quad 
{\cal J}\left(\derpar{}{q^j}\right) = -\left(\derpar{}{p_j}\right). 
$$ 
Now, we can consider the polarization $\P$, spanned by $\left\{ \derpar{}{\bar z_j}\right\}$ 
and as a symplectic potentials we can chose 
\beq \label{spot}
\Theta = \frac{i}{4}(\bar z_j \d z_j - z_j \bar z_j) 
\eeq
or the adapted one 
$$ 
\theta = \frac{i}{2}\bar z_j\d z_j .
$$ 
The polarized sections are the holomorphic sections of the complex line bundle 
$\Tan^*Q \times \Complex$ 
and using the symplectic potential (\ref{spot}), the quantum operators acting on these sections are: 
\beq 
\Op_{z_j}=-2\hbar\derpar{}{\bar z_j}+\frac{z_j}{2} 
\quad ; \quad 
\Op_{\bar z_j} \equiv \Op_{z_j}^+=2\hbar\derpar{}{z_j}+\frac{\bar z_j}{2}.
\label{opers} 
\eeq 
In physics this approach is considered more elegant reinforcing Dirac
notation, which depends upon the arguments of linear algebra. The raising and lowering operators (\ref{opers}),
or ladder operators, are the predecessors of the creation and annihilation operators used in the quantum mechanical description of interacting photons. The ladder operators obey commutation rules $\left [ \Op_{z_j},\, \Op_{\bar z_j}\right]=1$.

In this representation, a relevant role is played by the so-called number operator: 
$$ 
\Op_{\bar z_j z_j} = 2\hbar \left( z_j\derpar{}{z_j}-\bar z_j \derpar{}{\bar z_j} \right) .
$$ 
The polarized sections are: 
$$ 
\psi (z_j,\bar z_j) = F(z) e^{-\frac{z_j \bar z_j}{4\hbar}}, 
$$ 
which are holomorphic sections on $\Complex$, and the inner product is given by 
$$ 
\langle \psi_1 \mid \psi_2\rangle =\left( \frac{1}{2\pi \hbar}\right)^n 
\int_M F_1(z)\bar F_2(z) e^{-\frac{z_j\bar z_j}{2\hbar}}\Lambda_\Omega 
$$ 
Let us see the action of operators on polarized sections, which are the eigenstates of the system:
\begin{eqnarray} 
\Op_{z_j} \sta &=& z_j \psi = z_j F(z)e^{-\frac{z_j\bar z_j}{4\hbar}} 
\\ 
\Op_{z_j}^+\sta &=& 2\hbar\derpar{\psi}{z_j}+\frac{\bar z_j}{2}\psi 
= 2\hbar\derpar{F}{z_j}e^{-\frac{z_j\bar z_j}{4\hbar}} 
\\ 
\Op_{\bar z_j z_j}\sta &=& 2 \hbar\left( z_j\derpar{\psi}{z_j} -\bar 
z_j\derpar{\psi}{\bar z_j}\right) = \hbar z_j \derpar{F}{z_j} 
\end{eqnarray}
The eigenfunctions $F(z)$ are homogeneous polynomials of degree $n$, the action of number operator will reveals the twice of state number:
$$
Spec \left( \Op_{\bar z_j z_j}\right) = 2n.
$$
To obtain the correct spectrum, one should consider the proper Hamiltonian. There is a confusion to take the number operator instead of Hamilton operator and after introducing the complex coordinates, to write down the Hamiltonian as $\Op_H = \frac{1}{2}\Op_{\bar z_j z_j} $, which leads us to the incorrect physical spectrum. In these conditions a correct choice should be:
$$
\Op_H = \frac{1}{2}\left( \Op_{\bar z_j z_j} + \Op_{z_j \bar z_j} \right)
$$
and using the commutation relations $\left[ \Op_{ z_j} ,\, \Op_{\bar z_j} \right]=1$ we get:
\begin{equation}\label{opzHO}
\Op_H = \frac{1}{2} \left( \Op_{\bar z_j z_j} + 1 \right).
\end{equation}
one should get the correct spectrum:
\begin{equation}\label{sHO}
Spec \left( \Op_{\bar z_j z_j}\right) = n+\frac{1}{2}.
\end{equation}

In the following we use this representation using the standard CQ notations:
\begin{itemize}
\item ladder up/down operators will be denoted $a:=\Op_{z_j}$ and $a^+:= \Op_{\bar z_j}$.
\item the number operator $\rm N:= \Op_{\bar z_j z_j} $.
\item the Hamiltonian $ H:=\Op_H$.
\end{itemize}
 
\section{Examples and conclusions} 
 
\subsection{Illustrative Examples }

Let us consider the Schr\"odinger equation:
\beq 
\left(-\p_q^2+V(q)-E\right)\phi(q)=0,\label{schro3}
\eeq
the
parameter $E$ is the energy. 
This equation is the eigenvalue problem for the Hamilton operator.
In accordance with proposition \ref{ham}, to describe the HO states the Hamiltonian vector field have to be generated by:
\beq H:=- \p_q^2+\omega^2 q^2, \label{schro}
\eeq
where the momentum operator is written in its differential form (\ref{oppq}), we use natural units $\hbar=c=1$ and the parameter $\omega$ is proportional with the angular velocity.

Affine transformations, as translations $q\mapsto aq +b$ with
$a\neq0,$ $b$ constants, preserve the class of potentials.
In this manner all equations 
\beq\label{THO}
\left(-\partial_q^2+\omega^2 q^2+\rho
q+\lambda\right)\phi(q)=0,
\eeq 
given by parameters $\rho,\,\lambda$, named translated harmonic oscillator (THO) equations are in fact HO equations.
All elementary transformations as multiplication by a function or/and a change of variables, not depending on
the equation parameters, will reduce the THO equation (\ref{THO}) to the HO equation (\ref{schro}). 
An exception is made when $\omega=0$ case in which equation (\ref{THO}) becomes the Airy equation, which can be reduced to a special hypergeometric equation by a transformation.

\begin{example}\label{ex1}
Consider the following class of potentials:
\begin{eqnarray}
\label{pot1}
V_{\omega,\lambda}(q) =
\omega^2\left(\frac{3q}{2}\right)^{\frac23}+
\lambda\left(\frac{2}{3q}\right)^{\frac23}-\frac{5}{36}\frac{1}{q^2}.
\end{eqnarray}

Let us show that Schr\"odinger equations (\ref{schro3}), having this potential, with the following coordinate transformation 
$z=\left(\frac{3q}{2}\right)^{\frac23}$
will lead us to the Schr\"odinger equations (\ref{THO}). 
It is easy to verify that with the transformation operator: $U = z^\frac{1}{4}$ we obtain successively:
\begin{eqnarray*}&&
U^{-1}\Op_H U\\
&=&z^{\frac14-1}\left(-\partial_z^2+\omega^2 z^2+\rho
z+\lambda\right)z^{-\frac14}\\
&=&-\frac1z\partial_z^2+\frac{1}{2z^2}\partial_z+\omega^2
z+\rho+\frac{\lambda}{z}
- 
\left(\frac{1}{4}+\frac{1}{4^2}\right)\frac{1}{z^3}\\
&=&-\partial_q^2+\omega^2\left(\frac{3q}{2}\right)^{\frac23}+\rho+
\lambda\left(\frac{2}{3q}\right)^{\frac23}-\left(\frac{1}{4}+\frac{1}{4^2}\right)
\left(\frac{2}{3q}\right)^2. \end{eqnarray*}
We notice this is the starting potential rescaled with the energy $-\rho$. 
The physical system described by eq. (\ref{schro3}) with the potential (\ref{pot1}) admits GQ steps used for HO and 
according with the standard procedure, presented in section \ref{BFR}, the spectra of energy operator is given by (\ref{sHO}), but it is shifted with $-\rho$.
$$
\Op_H = \frac{1}{2} \left( \Op_{\bar z_j z_j} + 1 \right) + \rho.
$$
One can verify that the both conditions of prop. \ref{HOcond} are fulfilled, this is Hamiltonian is proper and bounded below and Hamiltonian vector field has elliptic critical points. 
A closer look to the potential reveals that its first two terms are a Lennard-Jones' type potential \cite{LJ} plus the repulsive inverse square potential.
\end{example}

\begin{example}\label{ex2}
Consider the following class of potentials:
\begin{eqnarray*}
V_{\rho,\lambda}(q) =
\frac{\rho}{(2q)^{\frac12}}+
\frac{\lambda}{2q}-\frac{3}{16}\frac{1}{q^2}
.\end{eqnarray*}
In the same manner, with suitable coordinate transformation:
$z=(2r)^{\frac12},$ we show th equivalence of the Hamilton operators of equations (\ref{schro3}) (\ref{THO}), with the transformation operator: $U= z^{-\frac{1}{2}}$ successively we get:
\begin{eqnarray*}&&
U^{-1}\Op_H U\\
&=&z^{\frac12-2}\left(-\partial_z^2+\omega^2 z^2+\rho
z+\lambda\right)z^{-\frac12}\\
&=&-\frac{1}{z^2}\partial_z^2+\frac{1}{z^3}\partial_z+\omega^2
+\frac{\rho}{z}+\frac{\lambda}{z^2}
-\left(\frac{1}{2}+\frac{1}{2^2}\right)\frac{1}{z^4}\\
&=&-\partial_q^2+\omega^2+\frac{\rho}{(2q)^{\frac12}}+\frac{\lambda}{2q}-
\left(\frac{1}{2}+\frac{1}{2^2}\right)\frac{1}{2^2q^2}
. \end{eqnarray*}
What we observe now is that the rescaling factor of the energy $-\omega^2$. The GQ will follow according with the standard procedure with the spectra of energy operator $\Op_H $, given by (\ref{sHO}), shifted with $-\omega^2$.
\end{example}

\begin{example}\label{ex3}
Consider the following class of potentials:
\begin{equation}\label{pot}
V_{\lambda,\mu}^s(q)= (\lambda+\sum_{k=0}^s C_k q^{2k})\,\exp(-\mu q^2),
\end{equation}
where the coefficients have the following expression:
$
C_k=\frac{(\lambda+k)\mu^{k}}{k!}\,.
$
with $\lambda$, $\mu$ real parameters and $s$ a positive integer. 
For each fixed parameters, it can be observed that in a vicinity of origin the potential behaves like HO potential, with a good approximation we can write:
$$
V_{\lambda,\mu}^s(r)=\lambda+\mu^2 q^2 - {\cal O} (q^{2s}).
$$
and discard the last term, which is a progressive summation of powers o $q^{2s}$ with grater than $s>1$. However, the potential do not accomplish the property of being proper neither the Hamiltonian vector field is complete albeit having elliptic critical points. In quantum mechanics it is not possible to approximate a potential in any condition, it will not preserve the particle probability of localization. One can make approximations upon potential, just if the system is in interaction and therefore apply perturbation theory. In a previous work \cite{pgo} we have proven that the system is integrable, but not in the HO sense. We found that the energy spectra of Hamiltonian is is different from that of HO.
\end{example}

\subsection{Conclusions}

In this paper we investigate the mathematical properties of the potential as part of the Hamiltonian, which describe the physical systems, things looked less in the approaches of physics. We aimed to find criteria for potential function in order to get an integrable system according to both CQ and GQ as well.
The dynamics of the system are given by the stationary points of Hamiltonian, in this view we considered stationary points of great interest in our investigation.
Thus, it have been shown that in a neighborhood of stable stationary point the potentials, via Morse lemma, can be always locally represented as its quadratic part. This condition is not a sufficient one, as example \ref{ex3} shows.
Further, describing the physical system by the associate momentum map we defined its reduction to HO by the existence of an isomorphism between their momentum maps. 
With the help of Eliasson's theorem, we give a result, by theorem \ref{HOcond}, which states the conditions for momentum map to be isomorphic with momentum map of HO. These conditions are found to refer to the property of Hamiltonian and Hamiltonian vector field, accordingly with their stationary points.
In accordance with physical laws of steady states just elliptic stationary points have to be considered. The global conditions apply to each potential individually, each integrable potential function describing a real physical system seems to accomplish the conditions of giving a global action. Exception seems to be made for periodically-repeating potential. In this case the potential domain can be, mathematically, restricted to one stationary point. This is ensured by Bloch\textsc{\char13}s theorem, which states that: {\em the system is solvable on each repeating domain}. The energy eigenfunctions have a basis consisting entirely of Bloch wave energy eigenstates. This decomposition of potential on smaller domains are gained by Eliasson's theorem as well, which states that stationary points are linearizable. 

Based on this reasoning, two physical systems were presented, namely example \ref{ex1} and example \ref{ex2}. An isomorphism with HO momentum map were found and the reduced systems were quantified using GQ as was introduced in section \ref{BFR}. 

In example \ref{ex3} a system being very similar with HO if the behavior of Hamiltonian is considered in a vicinity of the origin. This example emphasize the importance of the condition that the Hamiltonian to be proper. Without this condition even if the system is integrable, see \cite{pgo}, it is no momentum map isomorphism to the HO. To apply GQ there is need another approach than that from section \ref{BFR}. We conclude that no approximations can be made upon a potential. 
There are some some misinterpretations about the possibility to calculate with less errors at least first energy level in the case of Taylor approximation to HO of Hamiltonian in the neighborhood of a stationary point.


\begin{thebibliography}{00} 

\bibitem{2} P. M. Morse, Phys. Rev. 34, 57–64 (1929). 
\bibitem{3} C. Eckart, Phys. Rev. 35 (11), 1303–1309 (1930). 
\bibitem{4} N. Rosen and P. M. Morse, Phys. Rev. 42, 210–217 (1932). 
\bibitem{5} G. Pöschl and E. Teller, Z. Phys. 83 (3-4), 143–151 (1933). 
\bibitem{6} M. F. Manning and N. Rosen, Phys. Rev. 44 (11), 951–954 (1933). 
\bibitem{7} R. D. Woods and D. S. Saxon, Phys. Rev. 95 (2), 577–578 (1954). 
\bibitem{8} F. L. Scarf, Phys. Rev.112, 1137 (1958). 
\bibitem{pgo} F. Iacob and M. Lute, J. Math. Phys. 56, 121501 (2015).
\bibitem{Simm} D. J. Simms, Geometric Quantization of the Energy Levels in the Kepler Problem, Symposia Mathematica, vol. XIV, INDAM, Rome, 1973, Academic Press, New York, 1974 .
\bibitem{AM} R. Abraham and J. Marsden, Foundations of Mechanics, 2nd ed., Addison-Wesley, Reading, MA, 1978. p. 425, 429 
\bibitem{Kon}B. Kostant, Lecture Notes in Mathematics, Vol. 170 (New York, N. Y., 1970), p. 87. 

\bibitem{K} B. Kostant, Lecture Notes in Physics, Vol.6, 237 (New York, N. Y., 1970). 
\bibitem{AK} L. Auslander and B. Kostant, Invent. Math., 14, 255 (1971). 
\bibitem{S} J. M. Souriau, Symp. Math.,14, 343-360 (1974) 

\bibitem{Kumm}M. Kummer, Commun. Math. Phys., 84, 133-152 (1982) 
\bibitem{Cord}B. Cordani, Commun. Math. Phys., 113, 4, 649-657 (1988) 
\bibitem{Gr} J. A. Isenberg , General Relativity and Gravitation, 13, 4, 301-306, 1981 
\bibitem{MM}K. B. Marathe and G. Martucci, Nuovo Cimento B,79, 1 (1984) 
\bibitem{Puta} M. Puta, Rep. Math. Phys., 3, 27, 299-304, (1989) 
\bibitem{GS}G. Gaeta and M. Spera , Lett. Math. Phys., 16, 189-197, (1988) 

\bibitem{I} M. Puta and F. Iacob, Ann. Univ. Timisoara XXXIX (2001) 385-390. 

\bibitem{fluegge} S. Fl\" ugge, Practical Quantum Mechanics, Springer, -2nd print (1994) 
\bibitem{AHP} J. Derezi\' nski, M. Wrochna, Ann. Henri Poincare, 12, 2, 397-418, 2011
\bibitem{milson} Milson R., E. Horozov, Trends in mathematical physics, (Knoxville, TN, 1998), 289-297, AMS/IP Stud. Adv. Math., 13, Amer. Math. Soc., Providence, RI, 1999
\bibitem{Fordy} A.P. Fordy, Quantum Super-Integrable Systems as Exactly Solvable Models., SIGMA 3, (2007), 025. 
\bibitem{Arnold-cp}V. I. Arnold. Critical points of smooth functions Proc, ICM (Vamcouver 1974), 1, 19-41, (1975) 
\bibitem{Gordon}W. B. Gordon, On the completeness of Hamiltonian vector fields, Proc. A.M.S. 
26(1970), 329-331 
\bibitem{puta-pp} M. Puta, preprint ESI 177 (1994) 
\bibitem{Will} J. Williamson, Amer. J. Math., 58 (1996) 141-163 
\bibitem{Eliasson} L.H. Eliasson, Comment. Math. Helv., 65, (1990) 4-35.\\ 
L.H. Eliasson, Hamiltonian systems with Poisson commuting integrals, PhD thesis, University of Stockholm, (1984) 
\bibitem{Vey} Y. Colin de Verdiere and J. Vey, 18 (1979) 283-293 
\bibitem{Strocc} F. Strocchi, {\em An Introduction to the Mathematical Structure of Quantum Mechanics. A Short Course
for Mathematicians.} Advanced Series in Mathematical Physics, Vol. 27, World Scientific (2005).
\bibitem{LJ} Lennard-Jones, J. E. (1924), "On the Determination of Molecular Fields", Proc. R. Soc. Lond. A 106 (738): 463-477, 
\end{thebibliography}
\end{document}